\documentclass{svproc}
\usepackage[utf8]{inputenc}

\usepackage{url}

\usepackage[sectionbib]{natbib}
\bibpunct{(}{)}{;}{a}{}{,}%
\usepackage{graphicx}
\usepackage{amsmath,amssymb}
\usepackage{multirow}
\usepackage{longtable}

\usepackage[b5paper]{geometry}
\usepackage{subcaption}
\begin{document}
	\mainmatter              
	\title{Determining parameters of ULXs with model grids of extended atmospheres}
	\titlerunning{Determining parameters of ULXs with model grids of extended atmospheres}  
	%
	\author{A.~Kostenkov\inst{1, 2}, A.~Vinokurov\inst{1}
	 \and Y.~Solovyeva\inst{1}}
	\authorrunning{A.~Kostenkov et al.} 
	%
	\institute{Special Astrophysical observatory, Nizhnii Arhyz, Russia
		\and
		Saint-Petersburg University, Saint-Petersburg, Russia}
	
	\maketitle

\begin{abstract}
	In current work we continued study of wind parameters of ultraluminous X-ray sources (ULXs) using model grids of
extended atmospheres. We present sets of models with temperatures from 18000 up to 56000 K as a line equivalent widths (EWs) and their ratios diagrams. 
The fundamental wind parameters for some spectrally studied ULXs were estimated using EW diagrams. Also, influence of the wind velocity law on the equivalent widths of observed emission lines was investigated. {\it This study was funded by RFBR according to the research project 18-32-20214.}
\end{abstract}

%
Strong radiative-driven winds are characteristic not only of most classes of massive stars, but also of binary systems with supercritical accretion, such as ultraluminous X-ray sources (ULXs). \citet{Fabrika2015} have shown that optical spectra of ULXs are similar to those of LBV and WNLh stars (late nitrogen-sequence WR stars with hydrogen lines). The wind parameters estimates can be obtained using line EWs diagrams of the corresponding grids of models. In this work we continued study of ULXs winds using EWs diagrams of hydrogen and helium lines started in \citet{Kostenkov2020}.

\begin{figure}[h!]
\centering
\begin{subfigure}{.50\linewidth}
  \centering
  \includegraphics[width=0.7\linewidth]{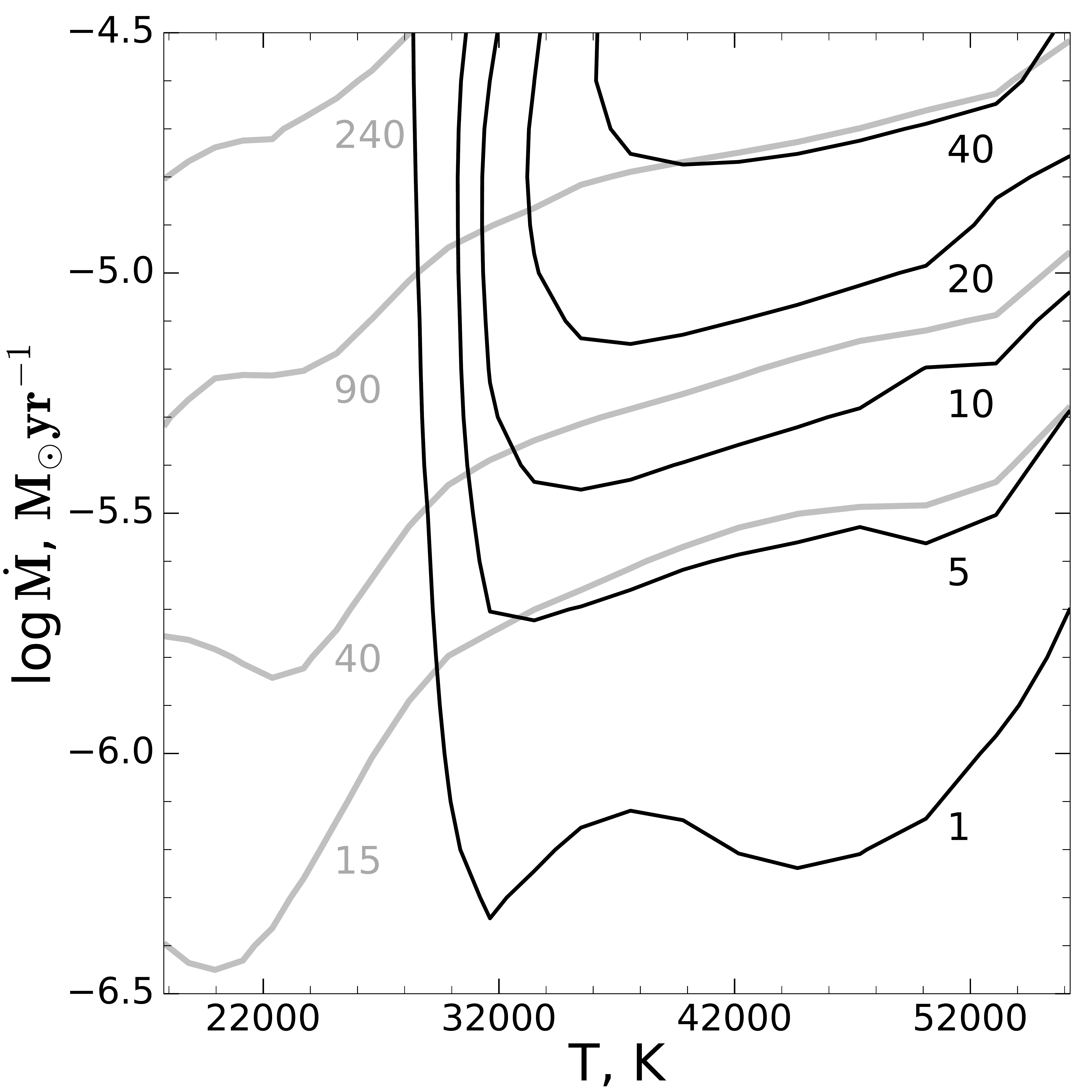}
  \label{fig:a}
\end{subfigure}%
\begin{subfigure}{.50\linewidth}
  \centering
  \includegraphics[width=0.7\linewidth]{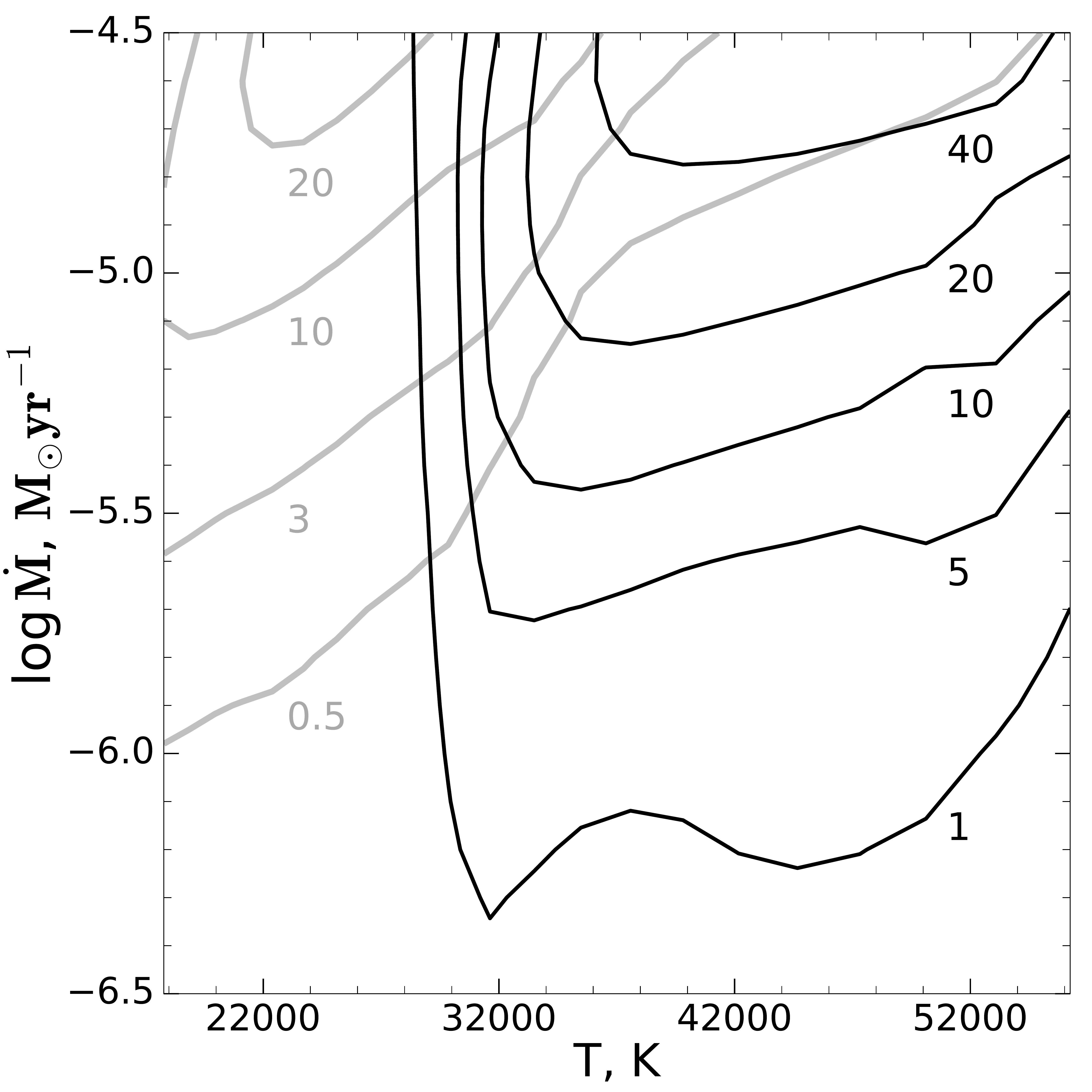}
  \label{fig:b}
\end{subfigure}
\caption{Equivalent widths diagrams of model grid with $\beta=1.5$ for ULXs for spectral resolution 5\AA{}; left: H$\alpha$ (grey lines) and He\,II $\lambda$4686 (black lines); right: He\,I $\lambda$5876 (grey lines) and He\,II $\lambda$4686 (black lines).}
\label{fig:ew_contour}
\end{figure}

\begin{figure}[h!]
	\centering
	\includegraphics[width=1.0\linewidth]{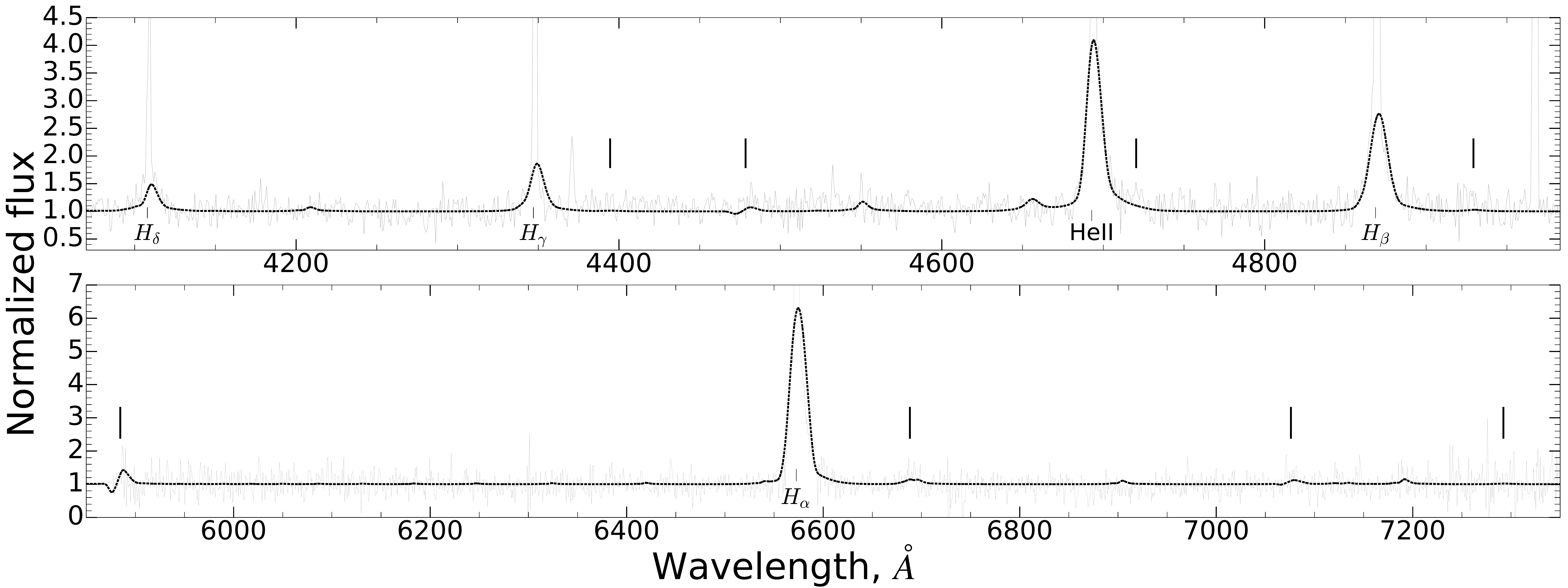}
	\caption{The observed spectrum of ULX in NGC~5408. (grey solid line; VLT/FORS2 archive data, the observation date is 2020 April 12) compared with the model spectrum smoothed with a spectral resolution of 5\AA {} (black solid line). He\,I lines are marked with black solid lines.}
	\label{ngc5408}
\end{figure}

Model were calculated with CMFGEN code \citep{Hillier1998} with temperatures $4.25 \leqslant \log{T_{*}} \leqslant 4.75$ (step 0.025), mass-loss rates ($M_{\odot}\text{yr}^{-1}$) $-6.5 \leqslant \log{\dot{M}} \leqslant -4.5$ (step 0.1) and wind velocity law $\beta=1.0,1.5$ with terminal velocity $V_{\infty}=800\,\text{km s}^{-1}$. Other parameters are similar to those used in \citet{Kostenkov2020}. EWs diagrams for $\beta=1.5$ are presented in Fig.\,\ref{fig:ew_contour}. Higher $\beta$ significantly enhance emission peaks, but also reduce lines widths. Difference in $\beta$ between our model grids mostly affect mass-loss rate estimates by $\log{\dot{M}} \approx 0.1$.
We determined wind parameters of NGC~5408 X-1 using presented above EWs diagrams. Best-fit of observed spectra was obtained with temperature $T_{*}=35480$\,K ($\log{T_*}=4.55$), mass-loss rate $\dot{M}=2\times10^{-5}\,M_{\odot}\text{yr}^{-1}$ and velocity law $\beta=1.5$. Spectrum of NGC~5408 X-1 with model is shown in Fig.\,\ref{ngc5408}.

\bibliographystyle{aa}
\bibliography{bibtexbase}

\begin{thebibliography}{3}
\expandafter\ifx\csname natexlab\endcsname\relax\def\natexlab#1{#1}\fi

\bibitem[{{Fabrika} {et~al.}(2015){Fabrika}, {Ueda}, {Vinokurov}, {Sholukhova},
  \& {Shidatsu}}]{Fabrika2015}
{Fabrika}, S., {Ueda}, Y., {Vinokurov}, A., {Sholukhova}, O., \& {Shidatsu}, M.
  2015, Nature Physics, 11, 551

\bibitem[{{Hillier} \& {Miller}(1998)}]{Hillier1998}
{Hillier}, D.~J. \& {Miller}, D.~L. 1998, \apj, 496, 407

\bibitem[{{Kostenkov} {et~al.}(2020){Kostenkov}, {Vinokurov}, {Solovyeva},
  {Atapin}, \& {Fabrika}}]{Kostenkov2020}
{Kostenkov}, A., {Vinokurov}, A., {Solovyeva}, Y., {Atapin}, K., \& {Fabrika},
  S. 2020, Astrophysical Bulletin, 75, 182

\end{thebibliography}
\end{document}